\documentclass[aps,twocolumn,showpacs]{revtex4}

\usepackage{epsf}
\usepackage{bm}

\begin{document}

\title{Single wall carbon nanotubes as coherent plasmon generators}

\author{I. V. Bondarev}

\affiliation{Department of Physics, North Carolina Central University, Durham, NC 27707, USA}

\begin{abstract}
The possibility of low-energy surface plasmon amplification by optically excited excitons in small-diameter single wall carbon nanotubes is theoretically demonstrated. The nonradiative exciton-plasmon energy transfer causes the buildup of the macroscopic population numbers of coherent localized surface plasmons associated with high-intensity coherent local fields formed at nanoscale throughout the nanotube surface. These strong local fields can be used in a variety of new optoelectronic applications of carbon nanotubes, including near-field nonlinear-optical probing and sensing, optical switching, enhanced electromagnetic absorption, and materials nanoscale modification.
\end{abstract}
\pacs{78.40.Ri, 73.22.-f, 73.63.Fg, 78.67.Ch}

\maketitle

\section{Introduction}

Single wall carbon nanotubes (CNs) --- graphene sheets rolled-up into cylinders of $\sim\!1\!-\!10$~nm in diameter and $\sim\!1$~$\mu$m up to $\sim\!1$~cm in length~\cite{Saito,Huang1,Huang2} --- are shown to be very useful as miniaturized electromechanical and chemical devices~\cite{Baughman}, scanning probe devices~\cite{Popescu1,Popescu2}, and nanomaterials for macroscopic composites~\cite{cncomposites1,cncomposites2,cncomposites3}.~The area of their potential applications was recently expanded to nanophotonics~\cite{Bond10jctn09oc1,Bond10jctn09oc2,Bond10jctn09oc3,Xia081,Strano,Hertel} after the demonstration of controllable single-atom incapsulation into single-walled CNs~\cite{Jeong}, and even to quantum cryptography since the experimental evidence was reported for quantum correlations in the photoluminescence spectra of individual nanotubes~\cite{Imamoglu}.

\begin{figure}[t]
\epsfxsize=7.5cm\centering{\epsfbox{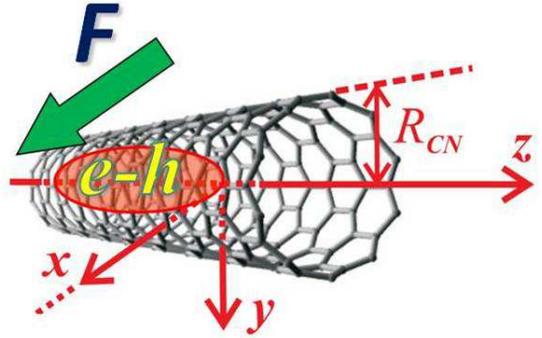}}
\caption{%
(Color online) The geometry of the problem.}
\label{fig1}
\end{figure}

The true potential of CN-based optoelectronic device applications lies in the ability to tune their properties in a precisely controllable way. In particular, optical properties of semiconducting CNs originate from excitons, and may be tuned by either electrostatic doping~\cite{Hertel,Spataru101,Spataru102}, or via the quantum confined Stark effect (QCSE) by means of an electrostatic field applied perpendicular to the nanotube axis~\cite{Bond09PRB,Bond11}.~In both cases the exciton properties are mediated by collective plasmon excitations in CNs~\cite{cnplasmons1,cnplasmons2}. In the latter case (Fig.~\ref{fig1}), we have shown recently that the QCSE allows one to control the exciton-interband-plasmon coupling in individual undoped CNs and their optical absorption properties, both linear~\cite{Bond09PRB} and nonlinear~\cite{Bond11}, accordingly.

In general, plasmons cannot be excited by light in optical absorption since they are longitudinal excitations while photons are transverse. In small-diameter ($\sim\!1$~nm) semiconducting CNs, light polarized along the CN axis excites excitons which, in turn, can couple to the nearest (same-band) interband plasmons~\cite{Bond10jctn09oc1,Bond10jctn09oc2,Bond09PRB}.~Both of these collective excitations originate from the same electronic transitions and, therefore, occur at the same energies $\sim\!1$~eV, as opposed to bulk semiconductors where they are separated by tens of eVs.~They do have different physical nature.~Their coexistence at the same energies in CNs is a unique general feature of confined quasi-1D systems where the transverse electronic motion is quantized to form 1D bands and the longitudinal one is continuous.

The formation of coupled exciton-plasmon excitations can be viewed as an additional non-radiative channel (in addition to phonons~\cite{Perebeinos05} and defects~\cite{Hagen05}) for the exciton relaxation in CNs, where optically excited excitons decay into low-energy interband plasmons.~In so doing, excitons generate the quanta of plasma oscillations on the CN surface, on the one hand, and this shortens their lifetime, on the other. Thus, by varying the exciton-plasmon coupling strength using the QCSE one controls both the radiative emission from an individual CN and surface electric field fluctuations associated with plasmons generated by excitons on the CN surface. This latter phenomenon is pretty much similar to the SPASER effect (Surface Plasmon Amplification by Stimulated Emission of Radiation) reported earlier for hybrid metal-semiconductor-dielectric nanosystems~\cite{Stockman}, and is the focus of the studies here for \emph{individual} small-diameter CNs.~The non-radiative exciton-to-plasmon energy transfer is shown to result in the large population numbers of coherent surface plasmons associated with high-intensity coherent oscillating fields concentrated at nanoscale across the CN diameter along the CN length. These strong local fields can be used in a variety of new CN based optoelectronic applications, including near-field nonlinear-optical probing and sensing, optical switching, enhanced electromagnetic absorption, and materials nanoscale modification.

\begin{figure}[t]
\epsfxsize=8.7cm\centering{\epsfbox{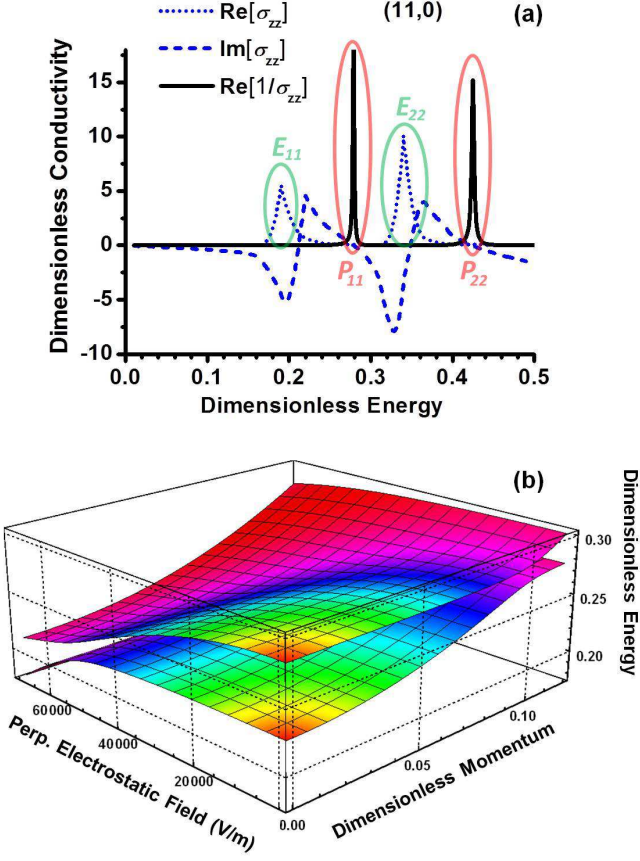}}
\caption{%
(Color online) Calculations for the (11,0) CN. (a)~Fragment of the energy dependence of the dimensionless (normalized by $e^2/2\pi\hbar$) axial surface conductivity $\sigma_{zz}$ along the CN axis~\cite{Bond09PRB}.~Ovals mark exciton ($E_{11}$, $E_{22}$) and interband plasmon ($P_{11}$, $P_{22}$) excitations [peaks of $\mbox{Re}\,\sigma_{zz}$ and $\mbox{Re}(1/\sigma_{zz})$, respectively]. (b)~Exciton-plasmon dispersion relation as a function of the perpendicular electrostatic field applied (see Fig.~\ref{fig1}) and longitudinal momentum for the lowest bright exciton [$E_{11}$ in~(a)] when coupled to the nearest interband plasmon [$P_{11}$ in~(a)]. Dimensionless momentum and dimensionless energy are defined as $3b[Momentum]/(2\pi\hbar)$ and $[Energy]/2\gamma_0$, respectively, where $b=1.42$~\AA~is the C-C interatomic distance and $\gamma_0=2.7$~eV is the C-C overlap integral.}
\label{fig2}
\end{figure}

\section{Plasmon generation by excitons}

In small-diameter semiconducting carbon nanotubes, because of their quasi-one-dimensionality, excitons are excited by the external electromagnetic (EM) radiation polarized along the CN axis~\cite{Ando}. As a consequence, they have their transition dipole moment and translational quasi-momentum both directed along the nanotube axis ($z$ axis of the problem --- see Fig.~\ref{fig1}; cylindrical coordinates are used).~That is why they are able to couple to their neighboring longitudinal interband plasmon modes~\cite{Bond10jctn09oc2,Bond09PRB}.~Figure~\ref{fig2} shows an example for the excitons and interband plasmons [Fig.\ref{fig2}(a)], and the exciton-plasmon dispersion relation [Fig.\ref{fig2}(b)] in the (11,0) CN (in part from Ref.\cite{Bond09PRB}, wherein similar data for other CNs are found).~In (b), we see the anti-crossing $\sim\!0.1$~eV, both in the energy-momentum plane and in the energy-field plane, revealing the QCSE as an efficient tool to control the exciton-plasmon coupling strength in individual CNs.

When the exciton is excited and the CN surface EM field subsystem is in the vacuum state, the time-dependent wave function of the whole system "exciton + surface EM field" is of the form
\begin{eqnarray}
|\psi(t)\rangle&=&\sum_{{\bf k},f}C_{f}(\textbf{k},t)\,e^{-i\tilde{E}_{f}({\bf k})t/\hbar} |\{1_f(\textbf{k})\}\rangle_{ex}|\{0\}\rangle\label{wfunc}\\
&+&\sum_{\bf k}\int_{0}^{\infty}\!\!\!\!\!\!d\omega\,C(\textbf{k},\omega,t)\,e^{-i\omega t}
|\{0\}\rangle_{ex}|\{1(\mathbf{k},\omega)\}\rangle.\nonumber
\end{eqnarray}
Here, $|\{1_f(\textbf{k})\}\rangle_{ex}$ is the single-quantum Fock state with one $f$-internal-state exciton excited of quasi-momentum ${\bf k}\!=\!\{k_\varphi,k_z\}$ with quantized $k_\varphi$ component (transverse quantization that results in electron-hole subbands) and continuous $k_z$ component representing the longitudinal motion of the exciton.~The excited single-quantum state $|\{1(\textbf{k},\omega)\}\rangle$ represents longitudinally polarized surface EM mode of frequency $\omega$ (the plasmon).~The respective vacuum states are $|\{0\}\rangle_{ex}$ for the exciton subsystem and $|\{0\}\rangle$ for the surface EM field subsystem. The coefficients $C_{f}(\textbf{k},t)$ and $C(\textbf{k},\omega,t)$ stand for the population probability amplitudes of the exciton subsystem and field subsystem, respectively.~The exciton energy is of the form
$\tilde{E}_{f}(\textbf{k})=E_f(\textbf{k})-i\hbar/\tau$, where $E_f(\mathbf{k})=E_{exc}^{(f)}(k_{\varphi})+\hbar^2k_z^2/2M_{ex}(k_{\varphi})$ stands for the total energy of the $f$-internal-state exciton and $\tau$ represents the phenomenological relaxation time to account for all possible (but into plasmons) slow exciton relaxation processes (normally attributed to the exciton-phonon scattering~\cite{Bond09PRB,Perebeinos05}). The energy $E_f(\textbf{k})$ consists of the exciton excitation energy $E_{exc}^{(f)}(k_{\varphi})=E_g(k_{\varphi})+E_b^{(f)}(k_{\varphi})$, where $E_g(k_{\varphi})=\varepsilon_e(k_{\varphi})+\varepsilon_h(k_{\varphi})$ is the band gap with $\varepsilon_{e,h}$ being the transversely quantized azimuthal electron-hole subbands, $E_b^{(f)}$ is the (negative) binding energy of the exciton, and the kinetic energy of the translational longitudinal motion of the exciton with the effective mass $M_{ex}=m_e+m_h$, where $m_{e}$ and $m_{h}$ are the (subband-dependent) electron and hole effective masses.

The time dependent Schr\"{o}dinger equation with the Hamiltonian of the whole system "exciton + surface EM field" applied on the wave function (\ref{wfunc}) results in the set of the coupled differential equations for $C_{f}(\textbf{k},t)$ and $C(\textbf{k},\omega,t)$ as follows~\cite{Bond09PRB}
\[
i\hbar\mbox{\it\.C}_f(\textbf{k},t)\,e^{-i\tilde{E}_{f}({\bf\!\,k})t/\hbar}\!=\!\int_{0}^{\infty}\!\!\!\!\!\!d\omega\,
\mbox{g}^{(+)}_f(\mathbf{k},\mathbf{k},\omega)\,C(\textbf{k},\omega,t)e^{-i\omega\!\,t}\!,
\]\vskip-0.5cm
\[
i\hbar\mbox{\it\.C}(\textbf{k},\omega,t)\,e^{-i\omega\!\,t}\!=\!
\sum_f\,[\mbox{g}^{(+)}_f(\mathbf{k},\mathbf{k},\omega)]^\ast
C_f(\textbf{k},t)e^{-i\tilde{E}_{f}({\bf\!\,k})t/\hbar}\!.
\]
Here, the interaction matrix element squared is given by
\begin{equation}
|\mbox{g}^{(+)}_f(\mathbf{k},\mathbf{k},\omega)|^2=\frac{\hbar S_0|d_{z}^f|^2\omega^3}{16\pi^3c^2R_{C\!N}^2}\,\mbox{Re}\frac{1}{\sigma_{zz}(\mathbf{k},\omega)}\,,
\end{equation}
where $d^f_z=\sum_{\mathbf{n}}\langle0|(\hat{\mathbf{d}}_\mathbf{n})_z|f\rangle$ with $\langle0|(\hat{\mathbf{d}}_\mathbf{n})_z|f\rangle$ being the electric dipole transition matrix element where the $f$-internal-state exciton is excited in the lattice site $\textbf{n}$ of the CN surface, $\sigma_{zz}$ stands for the CN surface axial conductivity with $\mbox{Re}(1/\sigma_{zz})$ representing the plasmon density of states [DOS, see Fig.~\ref{fig2}~(a)], $R_{C\!N}$ is the CN radius, and $S_0\!=\!(3\sqrt{3}/4)b^2$ is the area of an elementary equilateral triangle selected around each carbon atom in a way to cover the entire CN surface, $b\!=\!1.42$~\AA\space is the C-C interatomic distance.

In terms of the probability amplitudes $C_{f}(\textbf{k},t)$ and $C(\textbf{k},\omega,t)$, the exciton emission intensity distribution to generate plasmons is given by the final state probability at long times corresponding to the decay of all initially excited excitons, i.e., $I(\mathbf{k},\omega)=|C(\mathbf{k},\omega,t\!\rightarrow\!\infty)|^{2}\label{Iomega}$. This is related to the exciton probability amplitude $C_f(\mathbf{k},t)$, according to the equations above, and thereby is also associated with light absorption by excitons. The peak intensity represents the long-time plasmon population generated by optically excited excitons with the momentum~$\mathbf{k}$.

\section{Plasmon induced surface field}

Plasma oscillations generated by the non-radiative exciton decay into the nearest interband plasmon mode [Fig.~\ref{fig2}~(a)] can be viewed as \emph{standing} charge density waves due to the periodic opposite-phase displacements of the electron shells with respect to the ion cores in the neighboring elementary cells on the CN surface.~Such periodic displacements induce local coherent oscillating electric fields of zero mean, but non-zero mean-square magnitude, concentrated at the nanoscale across the diameter throughout the length of the nanotube.~The mean-square longitudinal local field magnitude can be calculated as the observable, the expectation value
\begin{equation}
E_z^2(\mathbf{n})\!=\!\langle[-\bm\nabla_{\!\mathbf{n}}\,\hat{\varphi}(\mathbf{n})]_z^2\rangle
\label{E2def}
\end{equation}
of the quantum electrodynamical (QED) longitudinal-electric-field operator at the lattice site $\mathbf{n}\!=\!\{R_{C\!N},\varphi_n,z_n\}$ of the CN surface in the state with the wave function (\ref{wfunc}). This results in (see Appendix A)
\begin{equation}
E_z^2(\mathbf{n})=\frac{\hbar L}{2\pi^2c^2R_{C\!N}}\sum_{\mathbf{k}=k_\varphi,k_z}f_{ex}(\mathbf{k},T)
\label{E2}
\end{equation}\vspace{-0.5cm}
\[
\times\int_0^\infty\!\!\!\!\!d\omega\,\omega^3\,\mbox{Re}\frac{1}{\sigma_{zz}(\mathbf{k},\omega)}\!\left[N(\mathbf{k},\omega)\!+\frac{1}{2}\right],
\]
where $N(\mathbf{k},\omega)=(4\pi c/L)I(\mathbf{k},\omega)$ is the number of plasmons generated by excitons with the momentum $\mathbf{k}$ of the 1st Brillouin zone on the surface of the tubule of length~$L$, and $f_{ex}(\mathbf{k},T)=\exp[-\hbar^2k_z^2/2M_{ex}(k_{\varphi})k_BT]/Q_{ex}$ is the exciton momentum distribution function with the partition function $Q_{ex}(T)=\sum_{\mathbf{k}}\exp[-\hbar^2k_z^2/2M_{ex}(k_{\varphi})k_BT]$.

Equation~(\ref{E2}) tells that in order to produce strong local mean-square fields, the exciton non-radiative emission resonance [given by $I(\mathbf{k},\omega\!\sim\!E_{exc})$] must overlap with the energy $E_p$ of the neighboring plasmon DOS resonance \{the peak of $\mbox{Re}[1/\sigma_{zz}(\omega\sim\!E_p)]$\}.~By way of example of the (11,0) CN in Fig.~\ref{fig2}, we see that this can be achieved by carefully tuning the exciton excitation energy $E_{exc}$ and the nearest interband plasmon energy $E_p$ by means of the QCSE. As this takes place, both $E_{exc}$ and $E_p$ experience red shift with increasing perpendicular electrostatic field.~However, the red shift of $E_{exc}$ is very small [barely seen in the constant momenta energy-field planes in Fig.~\ref{fig2}~(b)] due to the negative field-dependent contribution of the exciton binding energy $E_b$~\cite{Bond09PRB}.~As a consequence, $E_{exc}$ and $E_p$ [$E_{11}$ and $P_{11}$ in Fig.~\ref{fig2}~(a)] approach as the field increases, pushing the coupled exciton-plasmon system into the strong coupling regime where all of the optically excited excitons decay non-radiatively to generate interband (same-band) plasmons, yielding the peak optical absorption at the same time.~As the temperature grows up, higher momenta excitons start contributing to the process, lowering the field necessary to achieve the strong coupling regime [blue contrast in Fig.~\ref{fig2}~(b)].

\begin{figure}[t]
\epsfxsize=8.5cm\centering{\epsfbox{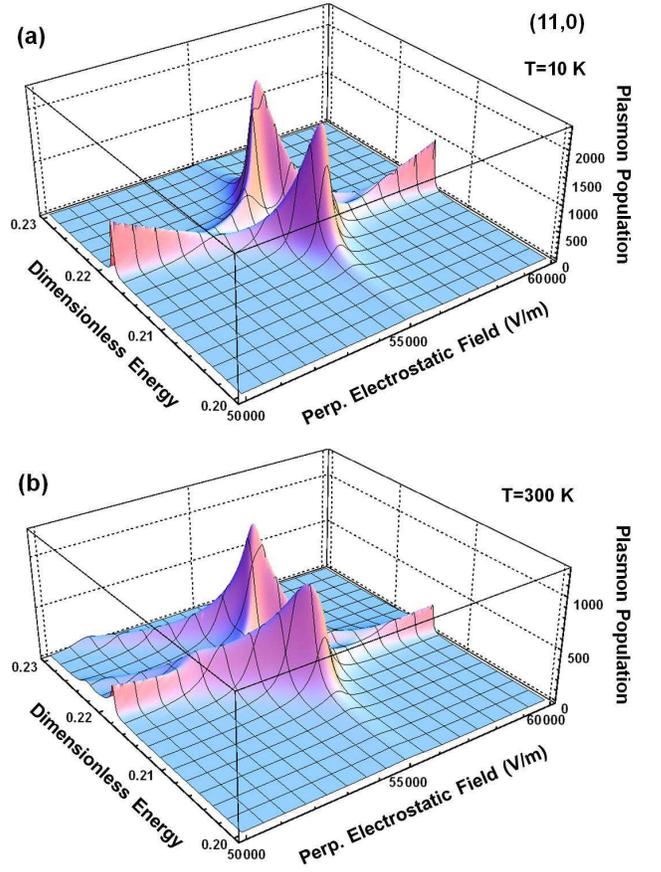}}
\caption{%
(Color online) Low (a) and high (b) longitudinal-momentum-averaged plasmon population (arb. units) for the non-radiative decay of the 1st bright exciton in the (11,0) CN. See Fig.~\ref{fig2} caption for the dimensionless energy.}
\label{fig3}
\end{figure}

Figure~\ref{fig3} shows plasmon population numbers averaged over the longitudinal momentum distribution~\cite{endnote1}, $\sum_{k_z}\!f_{ex}(\mathbf{k},T)N(\mathbf{k},\omega)$, also representing light absorption by excitons, calculated at low and high temperatures for the first bright exciton in the (11,0) CN with $E_{exc}$ and $E_p$ [$E_{11},P_{11}$ in Fig.~\ref{fig2}~(a)] tuned using the QCSE. The same exciton-plasmon parameters and their perpendicular-electrostatic-field dependences are used as in Ref.~\cite{Bond09PRB} ($E_{exc}\!=\!1.21$~eV, $E_b\!=\!-0.76$~eV, $E_{p}\!=\!1.51$~eV in the zero field,~and $\tau\!=\!\tau_{ph}\!=\!30$~fs to account for the exciton-phonon relaxation~\cite{Perebeinos05}).~We see the dramatic increase in the peak intensities, associated with increased optical absorption, when the field strength exceeds $5\times10^4$~V/m, both at low and at high temperatures~\cite{endnote2}. Rabi splitting occurs as the field drives the exciton-plasmon system into the strong coupling regime, whereby the effective plasmon generation starts. Temperature generally smoothens the effect due to higher momenta excitons contributing to the process.

\begin{figure}[t]
\epsfxsize=8.5cm\centering{\epsfbox{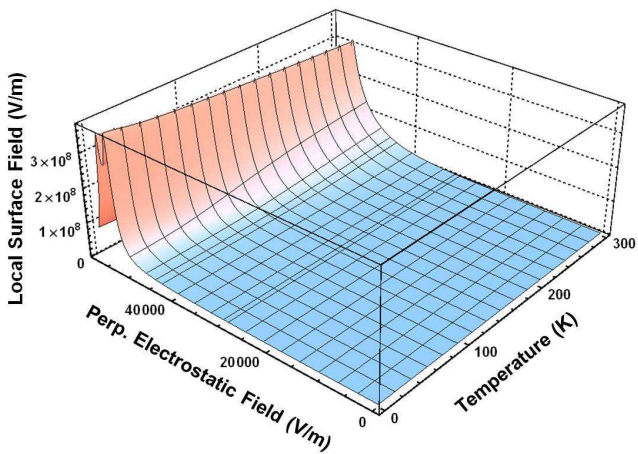}}
\caption{%
(Color online) Local surface field amplitude, $E_z$, as a~function of temperature and perpendicular electrostatic field applied, given by Eq.~(\ref{E2}) for plasmons generated by the non-radiative decay of the 1st bright exciton in the (11,0) CN.}
\label{fig4}
\end{figure}

Large plasmon population numbers are, according to Eq.~(\ref{E2}), associated with coherent oscillating fields concentrated locally throughout the CN surface. Figure~\ref{fig4} shows the calculations for the (11,0) CN. Huge local surface fields $\sim\!10^8$~V/m, just a few orders of magnitude less than intra-atomic fields, are created under the resonance conditions.~The effect slightly decreases with temperature, but it starts at lower electrostatic fields due to higher momenta excitons contributing to the plasmon generation.~Strong local surface fields created are the result of the efficient energy conversion, whereby the external EM radiation energy absorbed to excite excitons transfers into the energy of high-intensity coherent localized optical-frequency fields of charge plasma oscillations.

\section{Conclusions}

The effect presented here for \emph{individual} single wall CNs is analogous to the SPASER effect reported earlier for hybrid metal-semiconductor-dielectric nanostructures~\cite{Stockman}. The effect is universal as it originates from the transverse quantization of electronic degrees of freedom in quasi-1D systems.~It can manifest itself in densely packed aligned nanotube films as well, both through plasmon enhanced inter-tube Casimir interactions [$N\!=0\,$-term in Eq.~(\ref{E2})], as it is recently demonstrated for double wall CN systems~\cite{Popescu3}, and through the exciton-to-plasmon energy transfer tuned by means of the QCSE. In the latter case, plasmon-induced coherent local surface fields can be used in a variety of new tunable optoelectronic applications with both individual nanotubes and nanotube composites, such as enhanced electromagnetic absorption and optical switching, near-field nonlinear-optical probing and sensing, materials nanoscale modification.

\acknowledgments

Support from NSF (ECCS-1045661 \& HRD-0833184), NASA (NNX09AV07A), ARO (W911NF-11-1-0189), and DOE (DE-SC0007117) is acknowledged.

\appendix

\section{~Derivation of Eq.~(\ref{E2})}\label{appA}

Here, the exciton-plasmon coupling theory of Ref.~\cite{Bond09PRB}, developed earlier for individual single wall semiconducting CNs, is used to derive Eq.~(\ref{E2}) for the mean-square longitudinal local field induced by plasmons generated by optically excited excitons on the CN surface. In this theory, the nanotube is modelled by the neutral, infinitely long, infinitely thin, anisotropically conducting cylinder, where only the axial conductivity $\sigma_{zz}$ is taken into account while the azimuthal one is neglected being strongly suppressed by the transverse depolarization effect. The vector $\mathbf{e}_{z}$ of the orthonormal cylindrical basis $\{\mathbf{e}_{r},\mathbf{e}_{\varphi},\mathbf{e}_{z}\}$ is directed along the CN axis as shown in Fig.~\ref{fig1}.

The longitudinal surface electric field operator at the lattice site $\mathbf{n}=\{R_{C\!N},\varphi_n,z_n\}$ of the CN surface is of the form (Gaussian units)
\begin{equation}
-\bm{\nabla}_{\!\mathbf{n}}\,\hat{\varphi}(\mathbf{n})=\int_0^\infty\!\!\!\!\!d\omega\,
\hat{\underline{\mathbf{E}}}^\parallel(\mathbf{n},\omega)+h.c.,\label{phi}
\end{equation}
where
\begin{equation}
\underline{\hat{\mathbf{E}}}^{\parallel}(\mathbf{n},\omega)=
i\frac{4\pi\omega}{c^2}\sum_{\mathbf{m}}\,\!^{\parallel}\mathbf{G}(\mathbf{n},\mathbf{m},\omega)
\!\cdot\!\underline{\hat{\mathbf{J}}\!}\,(\mathbf{m},\omega)
\label{Enwpar}
\end{equation}
is the longitudinally polarized Fourier-domain surface electric field operator in the Schr\"{o}dinger picture with
$^{\parallel}\mathbf{G}(\mathbf{n},\mathbf{m},\omega)$ being the longitudinal part of the classical EM field Green tensor on the CN surface, and
\begin{equation}
\underline{\hat{\mathbf{J}}\!}\,(\mathbf{n},\omega)\!=\!\sqrt{\hbar\omega\,\mbox{Re}\,\sigma_{zz}(\omega)\over{\pi}}\,
\hat{f}(\mathbf{n},\omega)\textbf{e}_{z}
\label{currentCN}
\end{equation}
representing the CN axial surface current density operator defined in such a way as to preserve the fundamental QED equal-time commutation relations for the EM field operator components in the presence of the CN medium assisted absorption (see, e.g., Ref.~\cite{VogelWelsch}).~Here, $\hat{f}(\mathbf{n},\omega)$ along with its counter-part $\hat{f}^\dag(\mathbf{n},\omega)$ stand for the scalar bosonic field operators to annihilate and create, respectively, single-quantum EM field excitations of the frequency $\omega$ at the lattice site $\mathbf{n}$ of the CN surface.

The CN axial current density operator convoluted with the longitudinal EM field Green tensor per Eq.~(\ref{Enwpar}) yields the correctly quantized longitudinal surface electric field operator (\ref{phi}), whose $z$-component associated with the $zz$-component of the longitudinal Green tensor
\begin{equation}
^{\parallel}G_{zz}(\mathbf{n},\mathbf{m},\omega)=-\frac{c\,\sqrt{S_0}\,\delta(\varphi_n\!-\varphi_m)}
{8\pi R_{C\!N}\,\sigma_{zz}(\omega)}\;e^{i\omega|z_n-z_m|/c}\!,
\label{Gzz}
\end{equation}
is responsible for plasmon-induced coherent local surface fields as given by Eqs.~(\ref{E2def}) and (\ref{E2}).

The longitudinal Green tensor component (\ref{Gzz}) has the following property
\[
\sum_{\mathbf{n}}{^{\parallel}G_{zz}^\ast}(\mathbf{n},\mathbf{m},\omega){^{\parallel}G_{zz}}(\mathbf{n},\mathbf{m}^\prime\!,\omega^\prime)
\!=\!\!\left[2\pi\delta(\omega\!-\omega^\prime)\!-\!\frac{i}{\omega\!-\omega^\prime}\right]
\vspace{-0.5cm}
\]
\begin{equation}
\times\frac{c^3\delta(\varphi_m\!-\varphi_{m^\prime})\cos\!\left[\omega(z_m\!-z_{m^\prime})/c\right]}
{32\pi^2R_{C\!N}\sigma_{zz}^\ast(\omega)\sigma_{zz}(\omega^\prime)}\,.
\label{Gzz1st}
\end{equation}
This can be proven replacing the summation over lattice sites by the integration over the nanotube surface according to the rule
\[
\sum_\mathbf{n}\!\ldots=\frac{1}{S_0}\int\!d\mathbf{R}_n\!\ldots=
\frac{1}{S_0}\int_0^{2\pi}\!\!\!\!d\varphi_nR_{C\!N}\!\int_{-\infty}^\infty\!\!\!\!dz_n\!\ldots\,,
\]
where $S_0\!=\!(3\sqrt{3}/4)b^2$ ($b\!=\!1.42$~\AA\space being the C-C distance) is the equilateral triangle area selected around each carbon atom in such a way as to cover the entire CN surface. The integration over $\varphi_n$ is trivial, and the one over $z_n$ is performed by representing the cosine-function in the exponential form, writing the integral as follows
\[
\int_{-\infty}^\infty\!\!\!\!dz_n...=\lim_{L\rightarrow+\infty}\int_{-L/2}^{L/2}\!\!\!\!dz_n...
\]
with $L$ being the CN length, and dividing it into parts by means of the equation
\begin{eqnarray}
e^{i\omega|z_n-z_m|/c}=\theta(z_n\!-z_m)\,e^{i\omega(z_n-z_m)/c}\hskip0.3cm\nonumber\\[0.2cm]
+\,\theta(z_m\!-z_n)\,e^{-i\omega(z_n-z_m)/c}.\hskip1.cm\nonumber
\end{eqnarray}
This, after some algebra with the use of the following representation for the $\delta$-function (see, e.g., Ref.~\cite{Abers})
\begin{equation}
\delta(x)=\frac{i}{2\pi}\!\lim_{\begin{array}{c}\mbox{\scriptsize$\epsilon\!\rightarrow\!+0$}\\[-0.15cm]
\mbox{\scriptsize$t\!\rightarrow\!+\infty$}\end{array}}\!\frac{e^{-ixt}}{x+i\epsilon}\,,
\label{Abers}
\end{equation}
brings one to the equation
\[
\lim_{L\rightarrow+\infty}\int_{-L/2}^{+L/2}\!\!\!dz_n\,e^{-i\omega|z_n-z_m|/c+i\omega^\prime|z_n-z_{m^\prime}|/c}
\vspace{-0.35cm}
\]
\[
=2c\,\cos\!\left[\frac{\omega}{c}\left(z_m\!-z_{m^\prime}\right)\right]\!
\left[2\pi\delta(\omega\!-\omega^\prime)\!-\frac{i}{\omega\!-\omega^\prime}\right]\!
\]
to eventually result in Eq.~(\ref{Gzz1st}). The $\delta$-function representation (\ref{Abers}) can also be written in a more general form
\begin{equation}
\delta(x)=\frac{i^p(p-1)!}{2\pi\,t^{p-1}}\!\lim_{\begin{array}{c}\mbox{\scriptsize$\epsilon\!\rightarrow\!+0$}\\[-0.15cm]
\mbox{\scriptsize$t\!\rightarrow\!+\infty$}\end{array}}\!\!\frac{e^{-ixt}}{(x+i\epsilon)^p}\,,\;\;p=1,2,3,...
\label{Abersgen}\vspace{-0.25cm}
\end{equation}
which is proved by using the complex plane contour integration technique and the Cauchy integral theorem.

The property (\ref{Gzz1st}) can be converted to the momentum space to yield
\[
\sum_{\mathbf{n}}{^{\parallel}G_{zz}^\ast}(\mathbf{n},\mathbf{k},\omega)\,{^{\parallel}G_{zz}}(\mathbf{n},\mathbf{k}^\prime\!,\omega^\prime)=
\frac{Nc^3\,\delta_{\mathbf{k}\mathbf{k}^\prime}}{64\pi^3R_{C\!N}\,\sigma_{zz}^\ast(\omega)\sigma_{zz}(\omega^\prime)}
\vspace{-0.5cm}
\]
\begin{equation}
\times\left[2\pi\delta(\omega\!-\omega^\prime)\!-\frac{i}{\omega\!-\omega^\prime}\right],
\label{Gzz2st}
\end{equation}
where
\[
{^{\parallel}G_{zz}}(\mathbf{n},\mathbf{k},\omega)=
\frac{1}{\sqrt{N}}\sum_{\mathbf{m}}{^{\parallel}G_{zz}}(\mathbf{n},\mathbf{m},\omega)\,e^{i\mathbf{k}\cdot\mathbf{m}}
\]
with $N\!=2\pi R_{C\!N}L/S_{0}$ being the number of the lattice sites (carbon atoms) on the CN surface, and
\[
\delta_{\mathbf{k}\mathbf{k}^\prime}=\delta_{k_\varphi k_\varphi^\prime}\delta_{k_zk_z^\prime}
=\frac{1}{2\pi}\int_0^{2\pi}\!\!\!\!d\varphi_n\,e^{-i(k_\varphi-k_\varphi^\prime)\varphi_n}
\vspace{-0.35cm}
\]
\begin{equation}
\times\lim_{L\rightarrow+\infty}\frac{1}{L}\int_{-L/2}^{+L/2}\!\!\!dz_n\,e^{-i(k_z-k_z^\prime)z_n}.
\label{deltakkpr}
\end{equation}

From Eqs.~(\ref{phi})--(\ref{Gzz}), taking only the single-quantum states into account [in line with Eq.~(\ref{wfunc})], one has
\[
[-\bm\nabla_{\!\mathbf{n}}\,\hat{\varphi}(\mathbf{n})]_z^2=\int_{0}^{\infty}\!\!\!\!\!\!d\omega\,d\omega^\prime A^\ast(\omega)A(\omega^\prime)
\vspace{-0.25cm}
\]
\begin{equation}
\times\sum_{\mathbf{k},\mathbf{k}^\prime}{^{\parallel}G_{zz}^\ast}(\mathbf{n},\mathbf{k},\omega)\,
{^{\parallel}G_{zz}}(\mathbf{n},\mathbf{k}^\prime\!,\omega^\prime)\!
\label{nabla2}\vspace{-0.35cm}
\end{equation}
\[
\times\left[2\hat{f}^\dag(\mathbf{k},\omega)\hat{f}(\mathbf{k}^\prime\!,\omega^\prime)
+\delta_{\mathbf{k},\mathbf{k}^\prime}\delta(\omega-\omega^\prime)\right]\!,
\]
where
\begin{equation}
A(\omega)=i\frac{4\pi\omega}{c^2}\sqrt{\hbar\omega\,\mbox{Re}\,\sigma_{zz}(\omega)\over{\pi}}=-A^\ast(\omega)\,,
\label{Aomega}\vspace{-0.35cm}
\end{equation}
\[
\hat{f}^\dag(\mathbf{k},\omega)\!=\!\frac{1}{\sqrt{N}}\sum_{\mathbf{m}}\hat{f}^\dag(\mathbf{m},\omega)\,e^{i\mathbf{k}\cdot\mathbf{m}}\!,
\vspace{-0.5cm}
\]
\[
\hat{f}(\mathbf{k},\omega)\!=\!\left[\hat{f}^\dag(\mathbf{k},\omega)\right]^\dag\!\!,
\]
with the $\mathbf{k}$ summation running over the first Brillouin zone of the CN. Further, in view of the fact that all the lattice sites are equivalent, one has
\begin{equation}
[-\bm\nabla_{\!\mathbf{n}}\,\hat{\varphi}(\mathbf{n})]_z^2=\frac{1}{N}\sum_{\mathbf{m}}\,[-\bm\nabla_{\!\mathbf{m}}\,\hat{\varphi}(\mathbf{m})]_z^2
\label{1overN}
\end{equation}
for any arbitrary $\mathbf{n}$. Equations~(\ref{1overN}) and (\ref{Gzz2st}) allow one to reduce Eq.~(\ref{nabla2}) to the form
\begin{equation}
[-\bm\nabla_{\!\mathbf{n}}\,\hat{\varphi}(\mathbf{n})]_z^2=\int_{0}^{\infty}\!\!\!\!\!\!d\omega\,d\omega^\prime A^\ast(\omega)A(\omega^\prime)
\label{nabla2smplfd}\vspace{-0.35cm}
\end{equation}
\[
\times\,\frac{c^3}{32\pi^3R_{C\!N}\,\sigma_{zz}^\ast(\omega)\sigma_{zz}(\omega^\prime)}
\left[2\pi\delta(\omega\!-\omega^\prime)\!-\frac{i}{\omega\!-\omega^\prime}\right]
\vspace{-0.15cm}
\]
\[
\times\sum_{\mathbf{k}}\left[\hat{f}^\dag(\mathbf{k},\omega)\hat{f}(\mathbf{k},\omega^\prime)+\frac{1}{2}\,\delta(\omega-\omega^\prime)\right]\!.
\]

Now, the expectation value (\ref{E2def}) can be calculated. Equation~(\ref{E2def}) contains two types of averaging. They are the quantum mechanical averaging with the wave function (\ref{wfunc}) and the statistical one over the exciton quasi-momentum ${\bf k}$. Since the typical CN exciton excitation energies are $\gtrsim\!1$~eV (see, e.g., Ref.~\cite{Ma}), the excitons are assumed to be originally excited by the external EM radiation.~Once excited, the thermal exciton momentum relaxation starts in subbands. This situation is properly accounted for by the use of the density operator
\begin{equation}
\hat{\rho}(t)=\sum_{\mathbf{k}}f_{ex}(\mathbf{k},T)\;|\psi(\textbf{k},t)\rangle\langle\psi(\textbf{k},t)|\,,
\label{rho}
\end{equation}
where ${\bf k}\!=\!\{k_\varphi,k_z\}$ runs over the first Brillouin zone of the nanotube,
\[
|\psi(\textbf{k},t)\rangle=\sum_{f}C_{f}(\textbf{k},t)\,e^{-iE_f({\bf k})t/\hbar} |\{1_f(\textbf{k})\}\rangle_{ex}|\{0\}\rangle
\vspace{-0.35cm}
\]
\begin{equation}
+\int_{0}^{\infty}\!\!\!\!\!\!d\omega\,C(\textbf{k},\omega,t)\,e^{-i\omega t}|\{0\}\rangle_{ex}|\{1(\mathbf{k},\omega)\}\rangle
\label{wf}
\end{equation}
with the normalizing condition $\langle\psi(\textbf{k},t)|\psi(\textbf{k}^\prime,t)\rangle=\delta_{\textbf{k}\textbf{k}^\prime}$ [cf. Eq.~(\ref{wfunc})], and
\begin{equation}
f_{ex}(\mathbf{k},T)=\frac{1}{Q_{ex}}\exp\!\left[-\frac{\hbar^2k_z^2}{2M_{ex}(k_{\varphi})k_BT}\right]
\label{Boltzman}
\end{equation}
is the normalized classical (Boltzmann) exciton momentum distribution function with the partition function
\begin{equation}
Q_{ex}(T)=\sum_{\mathbf{k}}\exp\!\left[-\frac{\hbar^2k_z^2}{2M_{ex}(k_{\varphi})k_BT}\right]\!.
\label{Qex}
\end{equation}
This latter one can be evaluated explicitly as follows
\[
Q_{ex}(T)=\sum_{k_\varphi}\frac{L}{2\pi}\int_{-\tilde{k}_z}^{\tilde{k}_z}\!\!\!dk_z\exp\!\left[-\frac{\hbar^2k_z^2}{2M_{ex}(k_{\varphi})k_BT}\right]
\vspace{-0.25cm}
\]
\[
=L\sqrt{\frac{k_BT}{2\pi\hbar^2}}\sum_{k_\varphi}\sqrt{M_{ex}(k_{\varphi})}\;
\mbox{Erf}\!\left[{\sqrt\frac{\hbar^2\tilde{k}_z^2}{2M_{ex}(k_{\varphi})k_BT}}\,\right]
\vspace{-0.25cm}
\]
\[
\approx mL\sqrt{\frac{M_{ex}k_BT}{2\pi\hbar^2}}\,\mbox{Erf}\!\left(\!{\sqrt\frac{\hbar^2\tilde{k}_z^2}{2M_{ex}k_BT}}\,\right)\!.
\]
Here, the first Brillouin zone of the CN of $(m,n)$ type ($n\le m$) is taken to be consisting of $m$ parallel lines, as per quantized $k_{\varphi}=k_{\varphi}(s)=s/R_{C\!N}$ with $s=1,2,...,m$ and $R_{C\!N}=(\sqrt{3}\,b/2\pi)\sqrt{m^2+mn+n^2}$\,, each of length $2\tilde{k}_z=2B/k_{\varphi}(m)$, where $2B=2(4\pi^2/3\sqrt{3}\,b^2)$ is the rectangular area of the reciprocal space covered by the lines. This yields $\tilde{k}_z=2\pi/3b$ for the $(m,0)$ type CNs (zigzag) and $\tilde{k}_z=2\pi/\sqrt{3}\,b$ for the $(m,m)$ type CNs (armchair), in particular. The last expression above is the approximation neglecting the difference in the effective masses for different subband excitons, yielding $M_{ex}(k_{\varphi})\approx M_{ex}$ for all $k_{\varphi}$, to result in $\sum_{k_\varphi}\{...\}\approx\{...\}\sum_{s=1}^m\!1=\{...\}\,m$.

Using the density operator (\ref{rho}), one has
\begin{equation}
E_z^2(\mathbf{n})=\lim_{t\rightarrow+\infty}\mbox{Tr}\!\left\{[-\bm\nabla_{\!\mathbf{n}}\,\hat{\varphi}(\mathbf{n})]_z^2\,\hat{\rho}(t)\right\}\!.
\label{E2n}
\end{equation}
This expands further as follows
\[
E_z^2(\mathbf{n})=\lim_{t\rightarrow+\infty}\sum_{\mathbf{k}}\langle\psi(\textbf{k},t)|[-\bm\nabla_{\!\mathbf{n}}\,
\hat{\varphi}(\mathbf{n})]_z^2\,\hat{\rho}(t)|\psi(\textbf{k},t)\rangle
\vspace{-0.5cm}
\]
\[
=\lim_{t\rightarrow+\infty}\sum_{\mathbf{k}}f_{ex}(\mathbf{k},T)\langle\psi(\textbf{k},t)|
[-\bm\nabla_{\!\mathbf{n}}\,\hat{\varphi}(\mathbf{n})]_z^2\,|\psi(\textbf{k},t)\rangle,
\]
wherein, upon substitution of Eqs.~(\ref{nabla2smplfd}) and (\ref{wf}) in it, one has
\[
\langle\psi(\textbf{k},t)|\sum_{\mathbf{k}^\prime}\hat{f}^\dag(\mathbf{k}^\prime,\omega)\hat{f}(\mathbf{k}^\prime,\omega^\prime)
|\psi(\textbf{k},t)\rangle
\vspace{-0.5cm}
\]
\[
=C^\ast(\textbf{k},\omega,t)C(\textbf{k},\omega^\prime,t)\,e^{-i(\omega^\prime-\omega)t},
\]
to result in
\begin{equation}
E_z^2(\mathbf{n})=\lim_{t\rightarrow+\infty}\sum_{\mathbf{k}}f_{ex}(\mathbf{k},T)\!
\int_{0}^{\infty}\!\!\!\!\!\!d\omega\,d\omega^\prime A^\ast(\omega)A(\omega^\prime)
\label{Ez2final}\vspace{-0.35cm}
\end{equation}
\[
\times\,\frac{c^3}{32\pi^3R_{C\!N}\,\sigma_{zz}^\ast(\omega)\sigma_{zz}(\omega^\prime)}
\left[2\pi\delta(\omega\!-\omega^\prime)\!-\frac{i}{\omega\!-\omega^\prime}\right]
\vspace{-0.15cm}
\]
\[
\times\left[C^\ast(\textbf{k},\omega,t)C(\textbf{k},\omega^\prime,t)\,e^{-i(\omega^\prime-\omega)t}+\frac{1}{2}\,\delta(\omega-\omega^\prime)\right]\!.
\]
Multiplying the square brackets in here yields four terms to be taken in the limit of $t\rightarrow+\infty$, followed by the integration over $\omega^\prime$. The four terms are
\[
2\pi C^\ast(\textbf{k},\omega,t)C(\textbf{k},\omega^\prime,t)\,e^{-i(\omega^\prime-\omega)t}\delta(\omega\!-\omega^\prime)
\vspace{-0.4cm}
\]
\[
+\,iC^\ast(\textbf{k},\omega,t)C(\textbf{k},\omega^\prime,t)\frac{e^{-i(\omega^\prime-\omega)t}}{\omega^\prime-\omega}
\vspace{-0.4cm}
\]
\[
+\,\pi\delta(\omega\!-\omega^\prime)\delta(\omega\!-\omega^\prime)
\vspace{-0.4cm}
\]
\[
+\,\frac{1}{2}\,\delta(\omega\!-\omega^\prime)\frac{i}{\omega^\prime-\omega}\,.
\]
The first term here is trivial to deal with.~The second term is treated by formally adding a positive imaginary infinitesimal to its denominator to write $\omega^\prime\!-\omega+i\epsilon$ with $\epsilon\!\rightarrow\!+0$, followed by using Eq.~(\ref{Abers}). The term sums up with the first one. To treat the third term correctly, one first has to represent one of the $\delta$-functions as follows
\begin{equation}
\delta(\omega-\omega^\prime)=\lim_{L\rightarrow+\infty}\frac{1}{2\pi c}\int_{-L/2}^{+L/2}\!\!\!dz_n\,e^{-i(\omega-\omega^\prime)z_n/c}\,
\label{deltafunc}
\end{equation}
and then integrate the whole contribution over $\omega^\prime$. This gives the term proportional to the CN length $L$. Finally, the last, fourth term can be shown to yield zero principal value by noticing that it is an odd function of its variable. The $\delta$-function makes it zero for all $\omega\ne\omega^\prime$, where it must be zero, too, since it is odd. More accurate proof of this fact can be done in several ways, e.g., by using the identity $\delta^\prime(x)\!=\!-\delta(x)/x$ with $x\!=\!\omega-\omega^\prime$, followed by partial integration of the whole contribution, or starting from the $\delta$-function representation (\ref{deltafunc}), followed by manipulating the integration order over $\omega$ and $\omega^\prime$ with the use of the general properties of the integrands.

Inserting the contributions discussed into Eq.~(\ref{Ez2final}), adding them all together in the limit of $t\rightarrow+\infty$ with the use of Eq.~(\ref{Aomega}) for the coefficient $A(\omega)$, and taking into account the fact that $\mbox{Re}[1/\sigma_{zz}(\omega)]$ is a~sharp peak structure [where peaks should be attributed to different $k_{\varphi}(s)=s/R_{C\!N}$, $s=1,2,...$, to represent different interband transitions~\cite{endnote1}; see an example in Fig.~\ref{fig2}~(a)], one eventually arrives at the equation (\ref{E2}).


\begin{thebibliography}{99}
\bibitem{endnote1}$\sigma_{zz}(\mathbf{k},\omega)=\sigma_{zz}(k_\varphi,\omega)$ is assumed for the conductivity; weak longitudinal spatial dispersion is neglected~\cite{Bond09PRB}.
\end{thebibliography}
\end{document}